\documentclass[9pt,twocolumn,twoside]{pnas-new}

\templatetype{pnasresearcharticle} 

\newcommand{\pdf}{\mathrm{Pr}}

\title{An Objective Bayesian Analysis of Life's Early Start and Our Late Arrival}

\author[a,b]{David Kipping}
\affil[a]{Columbia University, Dept. of Astronomy, New York, NY 10027}
\affil[b]{Flatiron Institute, Computational Center for Astrophysics, New York, NY 10010}

\leadauthor{Kipping} 

\significancestatement{
Does life’s early emergence mean that it would re-appear quickly if
we were to re-run Earth’s clock? If the timescale for intelligence evolution
is very slow, then a quick start to life is actually necessary for our
existence - and thus doesn’t necessarily mean it’s a generally quick process.
Employing objective Bayesianism and a uniform-rate process assumption,
we use just the chronology of life’s appearance in the fossil record, that of
ourselves, and Earth’s habitability window, to infer the true underlying rates
accounting for this subtle selection effect. Our results find betting odds of
>3:1 that abiogenesis is indeed a rapid process versus a slow and rare
scenario, but 3:2 odds that intelligence may be rare.
}


\begin{abstract}
Life emerged on the Earth within the first quintile of its habitable window,
but a technological civilization did not blossom until its last. Efforts to
infer the rate of abiogenesis, based on its early emergence, are frustrated
by the selection effect that if the evolution of intelligence is a slow
process, then life’s early start may simply be a prerequisite to our
existence, rather than useful evidence for optimism. In this work, we interpret
the chronology of these two events in a Bayesian framework, extending upon
previous work by considering that the evolutionary
timescale is itself an unknown that needs to be jointly inferred,
rather than fiducially set. We further adopt an objective Bayesian approach,
such that our results would be agreed upon even by those using wildly
different priors for the rates of abiogenesis and evolution - common points of
contention for this problem. It is then shown that the earliest microfossil
evidence for life indicates that the rate of abiogenesis is at least 2.8
times more likely to be a typically rapid process, rather than a slow one.
This modest limiting Bayes factor rises to 8.7 if we accept the more disputed evidence of
$^{13}C$ depleted zircon deposits (Bell et al. 2015). For intelligence
evolution, it is found that a rare-intelligence scenario is slightly favored
at 3:2 betting odds. Thus, if we re-ran Earth’s clock, one should statistically favor life
to frequently re-emerge, but intelligence may not be as inevitable.
\end{abstract}

\dates{Competing interest statement: The authors declare no competing interests.}
\doi{\url{www.pnas.org/cgi/doi/10.1073/pnas.XXXXXXXXXX}}

\begin{document}

\maketitle
\thispagestyle{firststyle}
\ifthenelse{\boolean{shortarticle}}{\ifthenelse{\boolean{singlecolumn}}{\abscontentformatted}{\abscontent}}{}

\dropcap{A} fundamental question to modern science concerns the prevalence of
life, and intelligence, within the Universe. At the time of writing, searches
for non-terrestrial life within the Solar System have not yielded any direct
evidence for life \citep{mckay:1996,perchlorate:2010,waite:2017}, and the
remote detection of chemical biomarkers on extrasolar planets remains
years ahead of present observational capabilities \citep{marais:2002,
schwieterman:2018,totton:2018, fujii:2018}. The search for intelligence,
through the signatures of their technology, may be detectable under certain
assumptions \citep{cocconi:1959,dyson:1960,wright:2016,quasite:2019} and
limited observational campaigns have been attempted \citep{wright:2018}.
However, the underlying assumptions make it challenging to use these null
results to directly constrain the prevalence of life or intelligence at this
time.

Despite having no observational data concerning non-terrestrial
life, we are in possession of stronger constraints when it comes to life on
Earth. Until this situation changes, inferences concerning the existence of
life elsewhere in the Universe must unfortunately rely heavily on this single
data point \citep{spiegel:2012}. Whilst a single data point is not ideal, it is
certainly not devoid of information either \citep{simpson:2016}. This is even
true when strong selection biases are in play, such as the fact our existence
is predicated on at least one previously successful abiogenesis event.

Problems such as these lend themselves to Bayesian analysis, where the biases
can be encoded into the inference framework. The 2012 seminal paper of Spiegel
\& Turner \citep{spiegel:2012} applied this to the question of
abiogenesis. In that work, the authors treat abiogenesis as a Poisson process,
which holds for systems with a discrete number of successes in a finite
interval. This is used to define a likelihood function that accounts for the
possibility of multiple successes and the selection effect that a success is
demanded for us to observe ourselves. The authors conclude that the priors
ultimately dominate their posteriors, and that even choosing between three
reasonable and diffuse priors leads to greatly different answers. Accordingly,
it's very difficult to use the model to say anything definitive about how
difficult or easy abiogenesis really is.

Intuitively, the possibility that life typically emerges slowly seems highly
improbable given its relativity quick start on the Earth \citep{schopf:2006,
schopf:2007,bell:2015,schopf:2018}. Indeed, some commentators have remarked
that this fact implies that ``life is not a fussy, reluctant and unlikely
thing'' (ref. \cite{allwood:2016}, p. 501). The plausibility of the slow start scenario can
be understood to be a consequence of the selection effect, which requires that
life emerges fast enough for us to have sufficient time to evolve into complex
(``intelligent'') organisms capable of observing ourselves. The early emergence
of life on Earth then becomes consistent with a low abiogenesis rate, since
worlds where life does not emerge quickly never evolve to the point of an
intelligent observer.

This reveals the important role that the evolutionary timescale plays in
the inference problem, since it strongly shapes the selection bias effect.
In the Bayesian analysis of \cite{spiegel:2012}, the evolutionary timescale is
not known a-priori and thus is set to three different fiducial values (1, 2 and
3.1\,Gyr) in order to test the sensitivity of their results to this parameter.

In this work, we extend the model to allow the evolutionary timescale to itself
now become an inferred parameter. Rather than assume several fiducial values,
this parameter can be learnt by conditioning upon the time it took for
observers to evolve. This enables a joint posterior distribution between the
rate of abiogenesis and the rate of intelligence evolution that encodes the
covariance between the two. This not only yields a more robust estimate for
the abiogenesis rate by marginalizing over the uncertainty in the evolutionary
timescale, but it also infers the evolutionary timescale.

\section*{The Joint Likelihood Function}

\subsection*{Distribution of success times}

Following earlier work \cite{carter:2008,spiegel:2012,scharf:2016,chen:2018},
we describe abiogenesis as uniform rate (i.e. Poisson) process, defined by a
rate parameter $\lambda_L$. As with \cite{spiegel:2012}, we caution that this
does not imply that abiogenesis is a truly single step, instantaneous event.
Rather, we interpret the process to be a model which ``integrates out'' the
likely complex and multi-step chemistry which culminates in life. Indeed, it
may be that a variety of pathways can lead to abiogenesis, but this ensemble
is grouped into a single process where any of them succeeding counts as a
success for the ensemble. Further, it is not necessary to strictly define what
is meant by ``life'' here, only that the success of this Poisson process
ultimately led to the geological evidence for life, and that without it said
evidence would not exist.

The assumption of a uniform-rate process over some time interval can at first
seem problematic when one considers the stark changes to Earth's environment
over its history. However, much of this change is due to life affecting it's
environment and thus is a consequence of a success. The abiogenesis rate may
indeed be different after life begins, but it's also irrelevant since our
model only cares about the first success.

With these points in mind, we can now write that in a time interval
$t_L$, the probability of obtaining at least one successful abiogenesis event
($X_L>0$) will be

\begin{align}
\pdf(X_L>0;\lambda_L,t_L) =& 1 - \pdf(X_L=0;\lambda_L,t_L),\nonumber\\
\qquad =& 1 - e^{-\lambda_L t_L}.
\label{eqn:cumulative}
\end{align}

The time interval between successes for a Poisson process follows an
exponential distribution. This can be demonstrated by noting that
eq.~(\ref{eqn:cumulative}) corresponds to the probability of obtaining at
least one success over the interval of time $t_L$, and thus can be understood
as the cumulative probability distribution for the achievement of at least one
success by that time. Accordingly, the probability density function the first
success with respect to $t_L$ must be given by the derivative of
eq.~(\ref{eqn:cumulative}) with respect to time, yielding

\begin{align}
\pdf(t_L|\lambda_L) &= \lambda_L e^{-\lambda_L t_L}.
\end{align}

We now deviate from the approach of \cite{spiegel:2012} by considering a second
process, labelled by ``I'' for ``intelligence'' , which can only proceed once
the previous process (``L'') is successful.  The inverse of this process' rate
parameter, $\lambda_I$, describes the characteristic timescale it takes for
evolution to develop from the earliest forms of life, to an ``intelligent
observer'' which is carefully defined in the next subsection. We truncate the
times, such that both processes have to occur within a finite time $T$. Since
process I can only proceed once process L has occured, then this requires
$t_L+t_I<T$. The joint distribution of times $t_L$ and $t_I$ is therefore given
by

\begin{equation}
\pdf(t_L,t_I|\lambda_L,\lambda_I) \propto
\begin{cases}
\lambda_L \lambda_I e^{-\lambda_L t_L - \lambda_I t_I} & \text{if } t_L+t_I<T ,\\
0 & \text{otherwise } .
\end{cases}
\end{equation}

Imposing the condition of $t_L+t_I<T$ serves to truncate the joint distribution
and thus the above is formally a proportionality because it is not yet
normalized. After normalization, the expression becomes

\begin{equation}
\pdf(t_L,t_I|\lambda_L,\lambda_I) =
\begin{cases}
\frac{ \lambda_L \lambda_I (\lambda_L-\lambda_I) e^{-\lambda_L t_L - \lambda_I t_I} }{ \lambda_L (1- e^{-\lambda_I T}) - \lambda_I (1- e^{-\lambda_L T}) } & \text{if } t_L+t_I<T ,\\
0 & \text{otherwise } .
\end{cases}
\label{eqn:bivariatepdf}
\end{equation}

\subsection*{Accounting for observational constraints}

Before we discuss how eqn.~\ref{eqn:bivariatepdf} can be updated to
incorporated to include observational constraints, it is first useful to define
exactly what we mean by ``intelligence'' in this work. We adopt a functional
view of this term, consider that a successful event from process I is defined
as some kind of transition - which occurs after abiogenesis - which is
fundamentally necessary in order for analyses such as the one presented here to
be possible. In other words, this type of analysis is only possible because
process I succeeded, and would be impossible if it failed.

Expounding upon this, we can consider that process I results in an
observer/entity/society capable of i) obtaining and dating geological evidence
pertaining to the early emergence of life ii) the ability to model the
future climatic conditions of their world such that the habitability window can
be estimated, and iii) interpret the ramifications of this information
regarding the underlying rates of abiogenesis and evolution. For the sake
of brevity we'll refer to such outcomes as intelligent observers in what follows.
Formally, these three conditions are not equivalent to a technological
civilization, but this author would argue that it's difficult to imagine how
these feats would be possible in absence of one.

In what follows, we will attribute process I to correspond to the emergence of
human civilization and thus will further assume that no previous Earth dwelling
entities/observers/societies have had the capacity to satisfy the three
conditions discussed above. This assumption would be invalidated if the 
``Silurian hypothesis'' of \cite{schmidt:2019} were confirmed, which considers
the possible existence of industrial civilizations pre-dating humanity
\citep{wright:2018b}, in which case we would certainly advocate revisiting the
calculations that will be described in this paper.

The emergence of human civilization could be defined in a variety of ways. Some
possible defining ``moments'' could be the appearance of hominids, use the
evolution of homo-sapiens, complex language, the neolithic revolution, or the
first radio transmissions into space. Whatever we use, this only shifts $t_I$
around by several million years at most. Since $t_I$ is of order of several
Gyr, these disagreements have negligible impact on our final results and thus
$t_I$ will be treated as a fixed quantity.

This is not true for the first transition, since it seems to have occurred
relatively quick and the uncertainty associated with it is comparable to the
actual timing. Further, it is not possible to accurately date the emergence of
life, since any life could (and indeed must) have begun prior to its
appearance in the geological record. Accordingly, the true date for the
emergence of life, $t_L$, must pre-date the actual observation, $t_L'$,
i.e. $t_L<t_L'$. The probability of this can be calculated through integration:

\begin{align}
\pdf(t_L<t_L',t_I|\lambda_L,\lambda_I) &= \int_{t_L=0}^{t_L'} \pdf(t_L,t_I|\lambda_L,\lambda_I) \mathrm{d}t_L.
\end{align}

Since $t_I$ is defined as the time \textit{since} $t_L$, then one cannot
directly measure this value either. However, we can state that its value is
somewhere between $t_I'$ and $t_I'+t_L'$, where $t_I'$ is the observed time
difference between the emergence of intelligence and the first evidence for
life. This allows us to write our final likelihood function as

\begin{align}
\mathcal{L} =& \pdf(t_L<t_L',t_I'<t_I<t_I'+t_L'|\lambda_L,\lambda_I) \nonumber\\
\qquad=& \int_{t_I=t_I'}^{t_I'+t_L'} \pdf(t_L<t_L',t_I|\lambda_L,\lambda_I) \mathrm{d}t_I.
\label{eqn:finallikefn}
\end{align}

\begin{figure}
\begin{center}
\includegraphics[width=8.4cm,angle=0,clip=true]{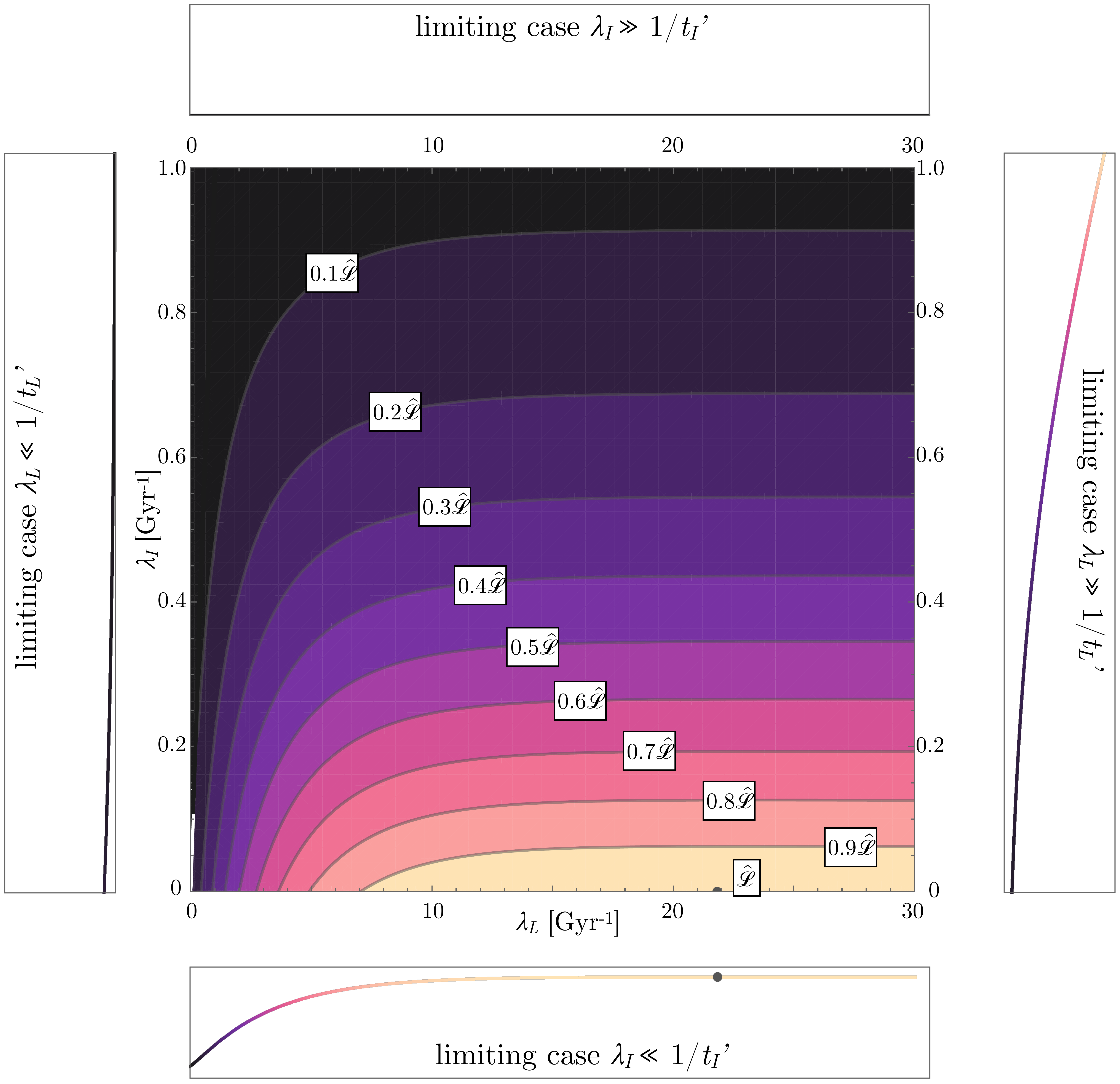}
\caption{
The joint likelihood function for the rate of abiogenesis, $\lambda_L$,
and the rate of intelligence emergence, $\lambda_I$ (for cases where
$T>2t_L'+t_I'$). Contours relative to the maximum likelihood position
(gray circle) are depicted by gray lines, where one can see a preference
against low $\lambda_L$ values. The limiting behaviors of the likelihood
function are shown along the edges of the figure.
}
\label{fig:likelihood}
\end{center}
\end{figure}

Evaluating the above yields a piecewise closed-form likelihood function
which is given in the SI Appendix. The function has two sub-domains, one which applies
to the interval $T>2t_L'+t_I'$, and one which applies to $T<2t_L'+t_I'$. As shown
later, the former case is applicable when $t_L'<0.904$\,Gyr and we plot this
function in fig.~\ref{fig:likelihood} against $\lambda_L$ and
$\lambda_I$, along with the limiting behaviors (see the SI Appendix).

It is interesting to note that the likelihood function is not monotonic and
has a global maximum which can be solved for numerically. For example, using
the ``optimistic'' data defined in the next section, it occurs at
$\hat{\lambda}_I\to0$ and $\hat{\lambda}_L=21.8$\,Gyr$^{-1}$ (the full
behavior is shown in the SI Appendix). However, along the $\lambda_I\to0$ axis, the
likelihood is almost flat beyond this peak. For example, with the same data,
the likelihood is 52.7\% of the peak when one sets
$\lambda_L = \hat{\lambda}_L/10$, but only 99.4\% of the peak when one sets
$\lambda_L = 10\hat{\lambda}_L$. Whilst maximum likelihood parameters are
instructive, we turn to Bayesian inference to determine the posterior
distributions and rigorously compare different model scenarios.

\subsection*{Adopted values for the observational data}

To perform any inference with our likelihood function, one first
needs to assign values to the observables $t_L'$ and $t_I'$, as well as $T$.
All three times are relative to some initial time when conditions on Earth
became suitable for life to emerge, and so let us first discuss how to define
this initial time.

There is of course uncertainty about the conditions on the early Earth and
when they became suitable for life \citep{sleep:2001,sleep:2006,arndt:2012}.
The Earth is generally thought to have been impacted by a Mars-sized body,
dubbed ``Theia'', 4.51\,Gyr ago in a cataclysmic event that formed the Moon
\citep{hartmann:1975}. Such an impact would have been a globally sterilizing
event, and indeed may have been accompanied by another sterilizing impactor,
``Moneta'', 40\,Myr later \citep{benner:2019}. Mineralogical
evidence from zircons indicates that both an atmosphere and liquid water
must have been present on the Earth's surface $(4.404\pm0.008)$\,Gyr ago
\citep{wilde:2001}. In this work, we consider these to be the necessary basic
requirements for abiogenesis to take place and thus adopt $T=4.408$\,Gyr
throughout. As noted by \cite{spiegel:2012}, a step-function like
transition from uninhabitable to habitable is surely too simplistic, but
our ignorance of Earth's early history and indeed the conditions necessary for
life mean that we do not at present have a well-motivated complex model to
impose in its place.

The earliest evidence for life arrives soon after this time, in the form of
$^{13}$C depleted carbon inclusions within 4.1\,Gyr old zircon deposits
\citep{bell:2015}. The source of this depletion remains controversial but
this would yield an optimistic estimate of $t_L'=0.304$\,Gyr. The earliest
direct and undisputed evidence for life comes from microfossils discovered in
3.465\,Gyr old rocks in western Australia \citep{schopf:2006,schopf:2007,
schopf:2018}, yielding $t_L'=0.939$\,Gyr. We highlight that, in both the
optimistic and conservative case, $t_L'$ is much larger than the 76\,Myr
uncertainty as to when the Earth became habitable and thus is not a dominant
source of uncertainty.

For $t_I'$, for reasons discussed in the previous subsection we can simply
attribute this time to be the modern era. We therefore adopt intelligent
observing as arriving ``now'' such that $t_I'=4.404\,\mathrm{Gyr}-t_L'$
in what follows.

Finally, we turn to $T$, which defines the interval over which the Earth
is expected to persist as habitable. It's important that we refine this
definition to be habitable for intelligent beings, such as ourselves. If
the Earth evolves into a state where only simple microbial life is possible,
then a success can no longer occur for process I. As the Sun evolves, its
luminosity will increase, which in turn increases the rate of weathering of
silicate rocks on the Earth \citep{malley:2013}. This increased weathering
draws down carbon from the atmosphere, thus gradually depleting the
atmospheric content of carbon dioxide. Once levels drop below $\sim$10\,ppm,
plant C4-photosynthesis will no longer be viable \citep{pearcy:1984},
leading to their imminent demise. It also possible that higher temperatures
and the progressive loss of Earth's oceans could also trigger an earlier die
off \citep{caldeira:1992}.

The end of plant life leads to a collapse in both the food chain and Earth's
oxygen productivity, upon which animal life is critically dependent. Large
endotherms, such as mammals and birds, will be the first to become extinct as a
result of their higher oxygen requirements \citep{falkowski:2005,
malley:2013}. Thus, one can reasonably consider that the habitable window for
intelligence decidedly ends once the reign of plant life comes to a close. The
timing for this is predicted to be 0.9\,Gyr by \cite{caldeira:1992}, a value
which is adopted in what follows to give $T=5.304$\,Gyr.

In summary, we set $T=5.304$\,Gyr but consider two values
for $t_L'$ of $t_L'=0.304$\,Gyr (``optimistic'') and $t_L'=0.939$\,Gyr
(``conservative''). This in turn gives two values of $t_I'$ of
$t_I'=4.404\,\mathrm{Gyr}-t_L'$.

\section*{The Non-Objective $\lambda$ Power-Law Prior}

\subsection*{The role of the prior}

Equipped with a likelihood function, one may infer the a-posteriori
distribution of $\lambda_L$ \& $\lambda_I$ using Bayes' theorem:

\begin{align}
\pdf(\lambda_L,\lambda_I|t_L',t_I') &= \frac{ \pdf(t_L',t_I'|\lambda_L,\lambda_I) \pdf(\lambda_L,\lambda_I) }{ \int \int \pdf(t_L',t_I'|\lambda_L,\lambda_I) \pdf(\lambda_L,\lambda_I) \,\mathrm{d}\lambda_L\,\mathrm{d}\lambda_I }.
\label{eqn:test1}
\end{align}

In any Bayesian inference problem, the posterior is a product of the likelihood
and the prior and thus is affected by both. In cases where one possesses
little or no information about the target parameters in advance, such as here,
the ideal prior should be minimally informative (``diffuse'') such that it
doesn't strongly influence the result \citep{jeffreys:1946}. In objective
Bayesianism, the resulting posterior should be expected to be universally be
agreed upon by everyone \citep{jaynes:1968} - whereas a subjective Bayesian
would argue that probability corresponds to the degree of personal belief 
\citep{finetti:1975}.

When equipped with strongly constraining data, even those using different
diffuse priors will generally find convergent solutions, since the likelihood
overwhelms the prior thus naturally leading to objective Bayesianism. This is
certainly not the case for our problem, since it has already been established
that the posterior for $\lambda_L$ is very sensitive to the priors
\citep{spiegel:2012}. In such a case, one should tread carefully and
seek a prior which can be objectively defined, such that other parties could
agree upon the choice of prior and thus the resulting posterior.

We therefore proceed by considering how to define a prior distribution which
is minimally informative and also not dependent upon subjective choices of the
prior distribution parameters. However, we will later show that several 
important inference statements can be made independent of the prior.

\subsection*{Power-law in $\boldsymbol{\lambda}$}

In previous work \citep{spiegel:2012,chen:2018}, the a-priori distribution for
$\lambda_L$ was assumed to be of the form $\lambda_L^n$ - a power-law. In
extending the likelihood to include $\lambda_I$, one may similarly extend this
power-law prior to encompass $\lambda_I$ by writing that
$\pdf(\lambda_L,\lambda_I) = \pdf(\lambda_L) \pdf(\lambda_I)$ (i.e. assuming
independence), where

\begin{equation}
\pdf(\lambda) =
\begin{cases}
\frac{\lambda^{-1}}{\log(\lambda_{\mathrm{max}})-\log(\lambda_{\mathrm{min}})} & \text{if } n=-1 ,\\
\frac{(n+1) \lambda^n}{\lambda_{\mathrm{max}}^{n+1}-\lambda_{\mathrm{min}}^{n+1}} & \text{otherwise }.
\end{cases}
\label{eqn:powerlaw}
\end{equation}

For $n=0$, this returns a uniform in $\lambda$ prior, for $n=-1$ a uniform in
$\log\lambda$ prior and for $n=-2$ a uniform in $\lambda^{-1}$ prior; the three
priors considered by \cite{spiegel:2012}. In adopting a prior of this form, one
must choose values for three shape parameters: the index, $n$ and the prior
bounds $\lambda_{\mathrm{min}}$ and $\lambda_{\mathrm{max}}$.

\subsection*{Assigning the prior shape parameters}

In \cite{spiegel:2012}, the favored index was $n=-1$ on the basis that a
log-uniform prior exhibits scale-invariance ignorance for $\lambda$. For
a real-valued parameter constrained only by a minimum and maximum threshold,
$n=-1$ also corresponds to the Jeffreys prior - a standard approach to
defining objective priors \citep{jeffreys:1946}. However, in this case, the
$n=-1$ power-law is not actually objective since one cannot objectively define
a minimum and maximum threshold. Because a power-law prior does not have
semi-infinite support from $\lambda=0$ to $\lambda=\infty$, then these prior
bounds have to be subjectively chosen.

For these bounds, \cite{spiegel:2012} set $\lambda_{\mathrm{max}} =
10^3$\,Gyr$^{-1}$ somewhat arbitrarily and a range of plausible values was
offered for $\lambda_{\mathrm{min}}$. Whilst useful as an exercise to test the
sensitivity of the posterior to the prior, this approach does not enable an
objectively defined solution. In an effort to objectively assign a power-law
prior, we consider here imposing the condition that the prior should be
\textit{fair} and \textit{unbiased}, which we will define in what follows.

As currently stated, the prior in eqn.~(\ref{eqn:powerlaw}) appears reasonably
diffuse for $n=0$, $-1$ and $-2$, and the bounds could essentially be
anything. However, it is worth recalling that the Poisson model is used as a
vehicle to describe the Bernoulli probability of one or more successful events
occurring. Accordingly, a natural alternative parameterization for this problem
is to consider the fraction of experiments in which the Poisson processes
culminate in at least one success, $f_L$ and $f_I$. Although these terms are
similar to the fractions defined in the Drake Equation \citep{drake:1965}, here
an ``experiment'' really refers to re-running Earth's history back and
observing how the stochastic processes play out each time (rather than some
other world). Since $f = 1 - e^{-\lambda T}$, then the power-law prior in
$\lambda$ is transformed into $f$-space as

\begin{align}
\pdf(f) &= \frac{n+1}{1-f} \frac{(-\log(1-f))^n}{ (-\log(1-f_{\mathrm{max}}))^{n+1} - (-\log(1-f_{\mathrm{min}}))^{n+1} }.
\label{eqn:9}
\end{align}

Recall that we seek to define a prior which is both fair and unbiased. For a
Bernoulli process, such as a coin-toss, a ``fair'' prior can be defined as one for
which the chance of a positively-loaded coin is no more or less likely than
negatively-loaded one. Accordingly, in our problem, we define a fair prior as
one which does not \textit{a-priori} favor either an optimistic ($f>1/2$) or
pessimistic ($f<1/2$) world-view. This can be quantified by defining the prior
odds ratio between the two scenarios using $\mathbb{F}$:

\begin{align}
\mathbb{F} &\equiv \frac{ \int_{f=1/2}^{f_{\mathrm{max}}} \pdf(f) \mathrm{d}f }{ \int_{f=f_{\mathrm{min}}}^{1/2} \pdf(f) \mathrm{d}f }.
\label{eqn:Fdefinition}
\end{align}

Setting $\mathbb{F}=1$, one may solve for $f_{\mathrm{max}}$ (which
corresponds to $\lambda_{\mathrm{max}}$) as a function of
$f_{\mathrm{min}}$ (corresponding to $\lambda_{\mathrm{min}}$):

\begin{equation}
\lim_{\mathbb{F} \to 1}\lambda_{\mathrm{max}} =
\begin{cases}
\frac{ (\log2)^2 }{ T^2 \lambda_{\mathrm{min}} }
 & \text{if } n=-1 ,\\
\frac{ \big( (2\log2)^{n+1} - ( \lambda_{\mathrm{min}} T )^{n+1} \big)^{1/(n+1)} }{T}
 & \text{otherwise }.
\end{cases}
\label{eqn:unbiasedlambdamax}
\end{equation}

Although this ensures a fair prior (subject to our definition), it does not
necessarily ensure an unbiased one. An unbiased prior is defined here as one for
which the \textit{a-priori} expectation value of $f$, given by $\textrm{E}[f]
\equiv \int_{f_{\mathrm{min}}}^{f_{\mathrm{max}}} f \pdf(f)\mathrm{d}f$,
equals one-half. After imposing the $\mathbb{F}=1$ constraint enabled by
eqn.~(\ref{eqn:unbiasedlambdamax}), we evaluate $E[f]$ in the limit of
$f_{\mathrm{min}} \to 0$ as a function of $n$. This reveals for $n=-1$,
the expectation value converges to a half, as desired for an unbiased prior
(see the SI Appendix). The only other value of $n$ in the range $-1<n<2$ that yields
a fair and unbiased distribution is $n=-0.709...$, but this is shown in the
SI Appendix to require an overly restrictive prior bound limit and thus is not used.

Using these two constraints, our fair and unbiased prior for $\lambda$ takes the
form $\lambda^{-1}$, with bounds following the relationship
given by eqn.~(\ref{eqn:unbiasedlambdamax}). Unfortunately, our two constraints
applied to three parameters is insufficient to uniquely define the prior - it
is still necessary to choose $\lambda_{\mathrm{min}}$ subjectively.
The posterior could still be argued to be objective if it were found to be
broadly insensitive to the choice of $\lambda_{\mathrm{min}}$ - however this
unfortunately turns out to be false, as shown in what follows.

\subsection*{Resulting posteriors}

To compute marginalized posteriors, we initially tried sampling using Markov Chain Monte Carlo and
nested sampling techniques, but found that the resulting posteriors were too
poorly sampled in the tails. Instead, we directly integrate the posterior
density in each parameter. For example, for the $\lambda_L$ marginalized
posterior, we slide along in a fine grid of $\lambda_L$ values and numerically
integrate (using the Gauss-Kronrod Rule) the joint posterior density over the
limits $\lambda_I=\lambda_{\mathrm{min}}$ to $\lambda_I=\lambda_{\mathrm{max}}$.

\begin{figure*}
\begin{center}
\includegraphics[width=18.0cm,angle=0,clip=true]{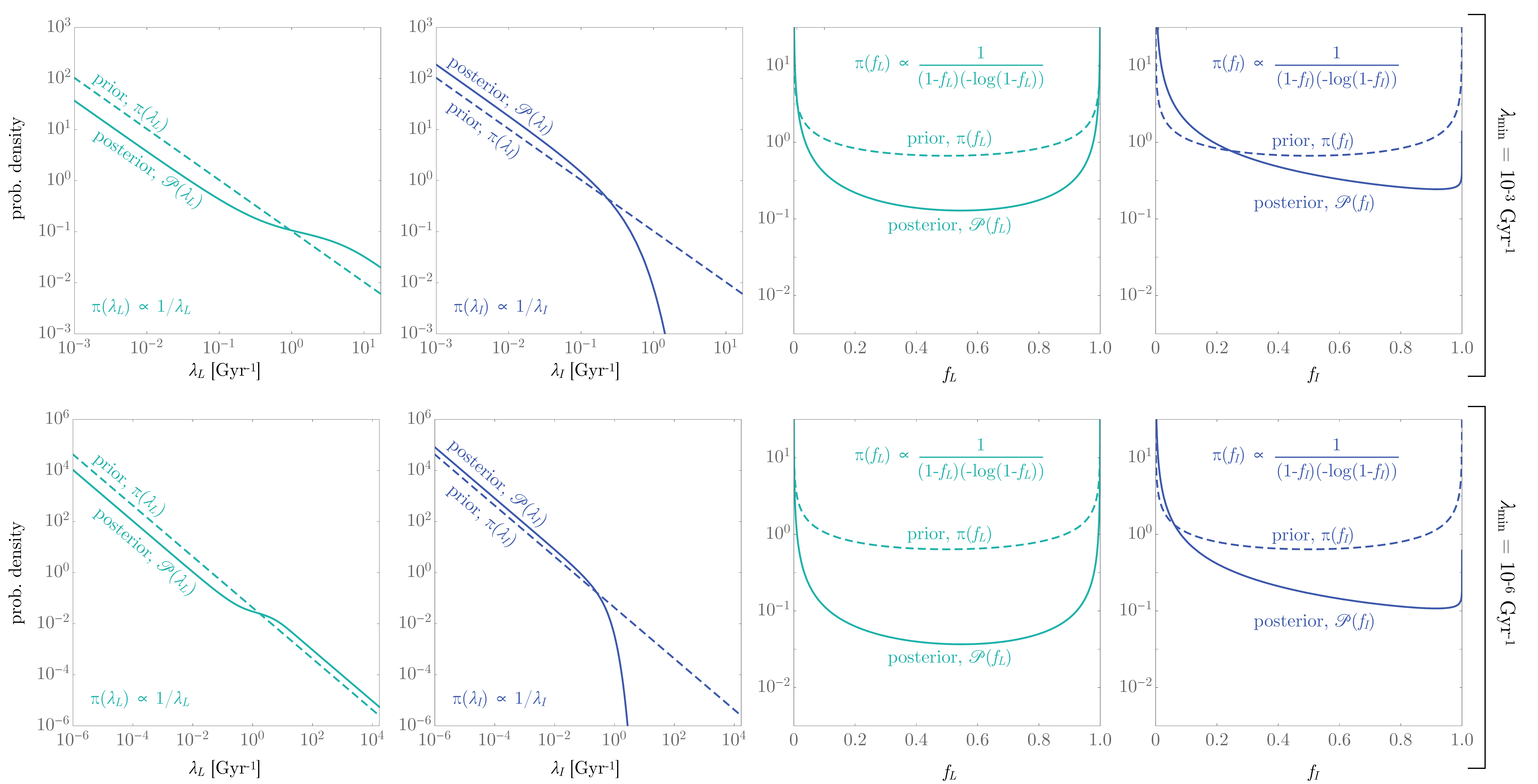}
\caption{
Marginalized posterior distribution (solid lines) for $\lambda_L$
(far-left), $\lambda_I$ (mid-left), $f_L$ (mid-right) and $f_I$
(far-right) using the log-uniform $\lambda$ prior (dashed lines) and
the optimistic data. Top-row assumes
$\lambda_{\mathrm{min}}=10^{-3}$\,Gyr$^{-1}$ and bottom-row assumes
$\lambda_{\mathrm{min}}=10^{-6}$\,Gyr$^{-1}$. The $y$-axis scale is chosen to
higlight the dynamic range.
}
\label{fig:lambda}
\end{center}
\end{figure*}

The resulting marginalized posteriors are shown in fig.~(\ref{fig:lambda}) for two
arbitrary choices of $\lambda_{\mathrm{min}}$: $10^{-3}$\,Gyr$^{-1}$ and
$10^{-6}$\,Gyr$^{-1}$. As can be seen from that figure, and perhaps not
surprisingly, the resulting distributions are certainly sensitive to the choice of
$\lambda_{\mathrm{min}}$. In conclusion, we argue here that a $\lambda$
power-law prior is simply unacceptable as an objective prior for this problem.

\section*{The Objective Bernoulli Prior}

\subsection*{Fair and unbiased priors with semi-infinite support}

Since we have no way of objectively choosing $\lambda_{\mathrm{min}}$,
then a fair and unbiased (subject to our definitions) power-law prior in
$\lambda$ does not provide a viable path to defining an objective posterior.

Rather than prescribing a prior in $\lambda$-space and then evaluating its
fairness and bias in $f$-space, we consider here simply writing down a fair and
unbiased prior in $f$-space directly. Since $f$ represents a fraction, it can
be interpreted as the Bernoulli probability of a success over the interval
$T$. For a Bernoulli process, a so-called Haldane prior \citep{haldane:1932}
of the form $\propto (f (1-f))^{-1}$ was argued by the objective Bayesian
Jaynes to represent the least informative prior \citep{jaynes:1968}. Such a
prior is fair and unbiased by construction, and places extreme weight on the
solutions $f=1$ and $f=0$ at the expense of intermediate values. The intuition
behind this is that either a very small fraction of planets will be successful
or almost all of them, but is unlikely the laws of nature are tuned such that
approximately half of the planets are successful.

Using the Fisher information matrix, one can define that the Jeffreys prior
for a Bernouilli distribution. The solution is not the Haldane prior, but
rather a softer variant of the form $\propto (f (1-f))^{-1/2}$. This translates
to a prior in $\lambda$ space given by

\begin{align}
\pdf(\lambda) &= \frac{T}{\pi \sqrt{e^{\lambda T}-1}}.
\end{align}

Both the Haldane and Jeffreys priors are fair and unbiased with respect to $f$
and indeed any prior of the more general form $\propto (f (1-f))^n$ satisfies
these conditions. However, the $n=-1$ case, corresponding to the Haldane prior,
is improper and indeed leads to an improper posterior too, but for all other
$n>-1$ the prior can be normalized to a finite quantity. Accordingly, the
Jeffreys prior is fair, unbiased, objectively defined and proper over the
interval $f=[0,1]$. Since $f=0$ corresponds to $\lambda=0$ and $f=1$
corresponds to $\lambda=\infty$, this naturally yields a proper prior with
semi-infinite support in $\lambda$, something which was not possible with the
power-law case discussed earlier. Together, these properties make the
distribution well-suited for our problem and we argue it solves the dilemma
faced in earlier work \citep{spiegel:2012}.

Although we consider the Jeffreys prior to be the ideal objective prior for
our problem, it is instructive to consider posteriors with $n=0$ (a uniform
prior in $f$) as well, which has a $\lambda$-space form of

\begin{align}
\pdf(\lambda) &= T e^{-\lambda T}.
\end{align}

\subsection*{Bayes factors independent of the $\lambda$ prior}

Equipped with our likelihood function and prior, one may now sample/integrate
the posterior probability distribution to compute marginalized distributions.
Marginalization irreversibly bakes the prior into the resulting collapsed
posteriors, but specific probabilistic statements can be made in a Bayesian
framework without marginalizing. In particular, we consider here an exercise
in Bayesian model comparison, where we seek to compare four models,
$\mathbf{\mathcal{M}}$, defined as the unique corners of the parameters volume:

\begin{itemize}
\item $\mathbf{\mathcal{M}_{00}}$: $\lambda_L \ll 1/t_L'$ and $\lambda_I \ll 1/t_I'$
\item $\mathbf{\mathcal{M}_{01}}$: $\lambda_L \ll 1/t_L'$ and $\lambda_I \gg 1/t_I'$
\item $\mathbf{\mathcal{M}_{10}}$: $\lambda_L \gg 1/t_L'$ and $\lambda_I \ll 1/t_I'$
\item $\mathbf{\mathcal{M}_{11}}$: $\lambda_L \gg 1/t_L'$ and $\lambda_I \gg 1/t_I'$
\end{itemize}

Binarizing the parameter volume into these four camps may at first seem
arbitrary - what about intermediate values? However, this partitioning is
consistent with objective Bayesianism. The objective Bernoulli prior treats
life/intelligence as being either very rare or very common, but unlikely to
be finely-tuned such that it approaches the intermediate value of one-half
- thus motivating the models above.

Conditioned upon some available data, $\mathcal{D}$, one may express that the
odds ratio between two models as $\pdf(\mathcal{M}_1|\mathcal{D})/
\pdf(\mathcal{M}_2|\mathcal{D}) = [\pdf(\mathcal{D}|\mathcal{M}_1)/
\pdf(\mathcal{D}|\mathcal{M}_2)] [\pdf(\mathcal{M}_1)/\pdf(\mathcal{M}_2)]$.
The terms inside the first square bracket is known as the Bayes factor, which
equals the odds factor under the simple assumption that no model is a-priori
preferred over any other. The Bayes factor is ratio of two ``evidences'' given
by $\mathcal{Z} \equiv \pdf(\mathcal{D}|\mathcal{M})$, and for the four models
defined above, $\mathcal{Z}$ can be expressed analytically and independent of
the prior $\pi(\lambda_L,\lambda_I)$. This can be seen by noting that, for
example with model $\mathcal{M}_{00}$:

\begin{align}
\mathcal{Z}_{00} =& \pdf(\overbrace{t_L',t_I'}^{\mathcal{D}}|\overbrace{\lambda_L \ll 1/t_L',\lambda_I \ll 1/t_I'}^{\mathcal{M}_{00}}) \nonumber\\
\qquad=& \lim_{\lambda_L\ll1/t_L'} \lim_{\lambda_I\ll1/t_I'} \underbrace{\pdf(t_L',t_I'|\lambda_L,\lambda_I)}_{=\mathcal{L}}
\end{align}

Thus, the evidences of these four corner models is independent of the
prior of $\pi(\lambda_L,\lambda_I)$, meaning that even those adopting
different priors would consistently agree on the Bayes factors. We note
that \cite{spiegel:2012} used a similar strategy and we provide
an alternative explanation of this prior-free model comparison in the
SI Appendix, in terms of the Savage-Dickey ratio \citep{dickey:1971}.

In practice, Bayes factors for the two corners with rapid intelligence
emergence ($\lambda_I \gg 1/t_I'$) tend to zero, since this is the behavior
of the likelihood function (see the SI Appendix). This can be understood by the fact
that if the rate of intelligence emergence were extremely fast, then it would
be incompatible with taking as long as it did here on Earth. In contrast, life
could emerge much faster than $t_L'$ because $t_L'$ only represents the first
appearance of life in the geological record, not the actual date of
abiogenesis.

Since these two corners are zero, we instead compare models $\mathcal{M}_{10}$
to $\mathcal{M}_{00}$. This represents the Bayes factor between a scenario
where abiogenesis is a fast versus a slow process conditional upon the
premise that intelligence emergence is itself a slow process. In this case, we
find

\begin{equation}
\frac{\mathcal{Z}_{10}}{\mathcal{Z}_{00}} =
\begin{cases}
\frac{T}{2t_L'}
 & \text{if } T>2t_L'+t_I ,\\
\frac{T t_L'}{4(T-t_I') t_L' - 2t_L' - (T-t_I')^2}
 & \text{if }T<2t_L'+t_I,
\end{cases}
\end{equation}

which evaluates to $(\mathcal{Z}_{10}/\mathcal{Z}_{00}) = 8.73$ and $2.83$ for
the optimistic and conservative data inputs respectively. The above also
reveals the dependency on $T$ is nearly linear; if $T$ is revised significantly
up then optimism for life would also increase. We highlight that
$\mathcal{Z}_{10}<\mathcal{Z}_{00}$ if $t_L'>3.72$\,Gyr - i.e. if the earliest
evidence for life were from no earlier than 680\,Myr ago, we would conclude
that abiogenesis was an improbable event.

\begin{figure}[t]
\begin{center}
\includegraphics[width=8.4cm,angle=0,clip=true]{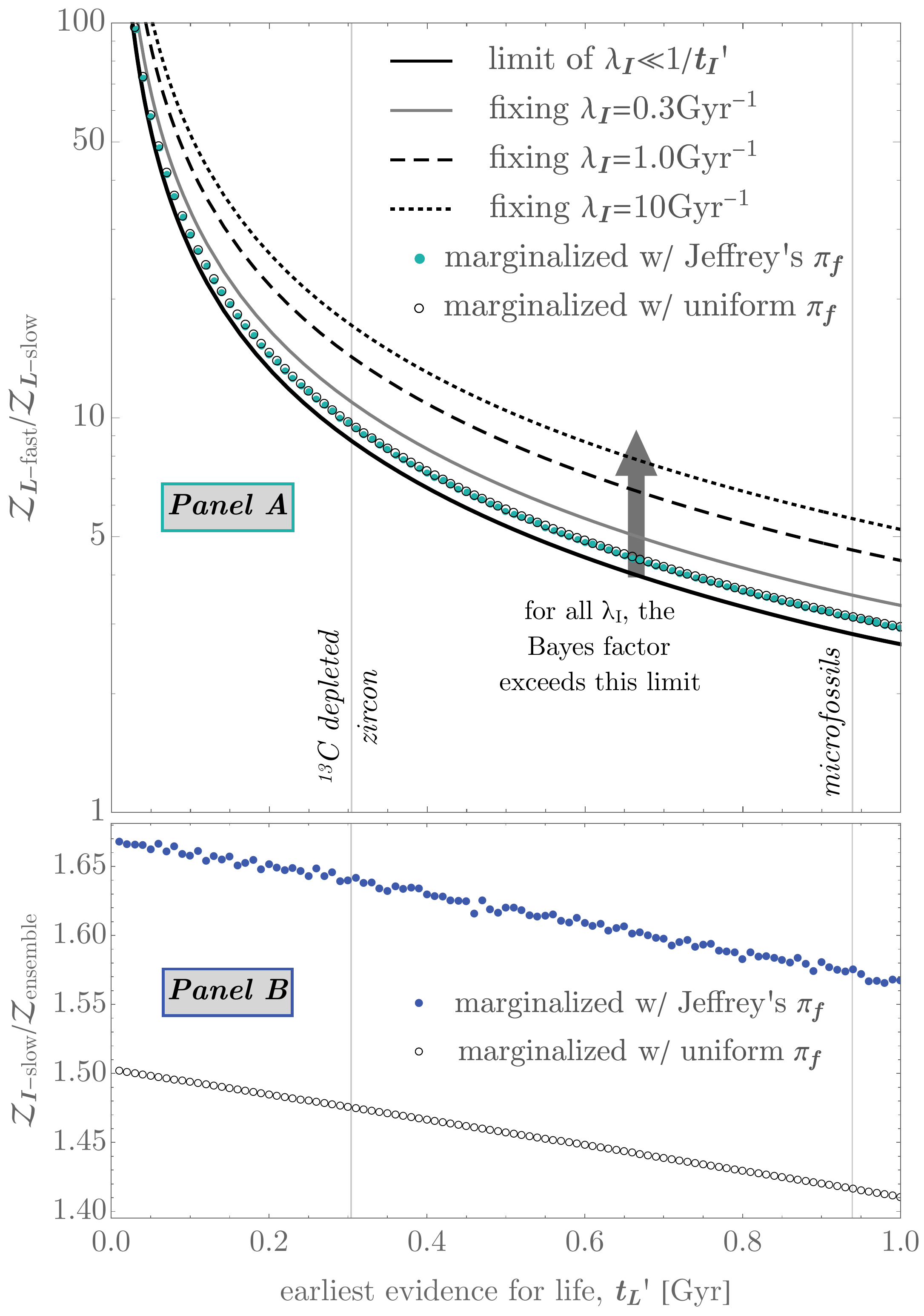}
\caption{
A: Bayes factor between a model where life emerges rapidly
($\lambda_L \gg 1/t_L'$) versus slowly ($\lambda_L \ll 1/t_L'$) on the
Earth. A quick start is favored by at least a factor of three conditioned
upon early microfossil evidence, independent of our assumptions
regarding the evolutionary timescale of intelligent observers and priors on
the abiogenesis rate. B: Bayes factor of a scenario where intelligent observers
typically emerge much longer than occurred on Earth, versus the ensemble
of possibilities. There is a weak preference for a rare intelligence scenario.
}
\label{fig:bayesfactor}
\end{center}
\end{figure}

If we relax the assumption that $\lambda_I \ll 1/t_I'$ and let intelligence
become faster, then the Bayes factor monotonically rises, as shown in
fig.~\ref{fig:bayesfactor}. This therefore means that the Bayes factor of a 
quick versus slow abiogenesis scenario must be greater than the
limiting case of $\mathcal{Z}_{10}/\mathcal{Z}_{00}$, irrespective of whatever
value $\lambda_I$ takes (or indeed whatever the prior is).

On this basis, we can conclude that even with the most conservative date
for the emergence of life, a scenario where abiogenesis occurs
rapidly is at least three times more likely than a slow emergence,
independent of the priors and even the timescale it takes for intelligence
to emerge. If the more ambiguous evidence for an earlier start to
life is confirmed \citep{bell:2015}, then this would increase the odds
to a factor of nine, representing relatively strong preference for a model
where life would consistently emerge rapidly on the Earth, if time were
replayed.

\subsection*{Bayes factors after marginalization}

Thus far we have avoided using the marginalized posteriors, which has the
benefit of enabling model comparison independent of the prior
$\pi(\lambda_L,\lambda_I)$. However, it also has the disadvantage that we
can only compare conditional scenarios. For example, our result for
$(\mathcal{Z}_{10}/\mathcal{Z}_{00})$ is a Bayes factor conditional upon
the assumption of a slow intelligence emergence. Whilst it turns out this
can be interpreted as a lower limit on a fast versus slow abiogenesis
scenario, marginalization allows for a calculation which integrates over
the uncertainty in $\lambda_I$. For example, one may write that the evidence
for a model where $\lambda_L\ll1/t_L'$ (slow abiogenesis) marginalized over
$\lambda_I$ is given by

\begin{align}
\mathcal{Z}_{\mathrm{L-slow}} &= \int_{\lambda_I=0}^{\infty} \Big(\lim_{\lambda_L\ll1/t_L'}\mathcal{L}\Big) \pi(\lambda_I) \mathrm{d}\lambda_I.
\end{align}

We numerically evaluated the evidences for $\mathcal{Z}_{\mathrm{L-slow}}$
using the above, $\mathcal{Z}_{\mathrm{L-slow}}$ ($\lambda_L\gg1/t_L'$) as well
as $\mathcal{Z}_{\mathrm{I-slow}}$ ($\lambda_I\ll1/t_I'$) and the ensemble
evidence over all possibilities, $\mathcal{Z}_{\mathrm{ensemble}}$. As before,
the fast intelligence emergence scenario has zero evidence since the likelihood
tends to zero in this regime, for reasons discussed earlier. The resulting
Bayes factors for
$(\mathcal{Z}_{\mathrm{L-fast}}/\mathcal{Z}_{\mathrm{L-slow}})$ are shown
in panel A of fig.~\ref{fig:bayesfactor} by the closed (Jeffreys prior) and
open (uniform prior) circles.

The two sets of points are almost indistinguishable and consistently lie above
our previously derived lower limit on the Bayes factor, as expected. Using the
optimistic data, we find that
$(\mathcal{Z}_{\mathrm{L-fast}}/\mathcal{Z}_{\mathrm{L-slow}}) = 9.538$ and
$(\mathcal{Z}_{\mathrm{L-fast}}/\mathcal{Z}_{\mathrm{L-slow}}) = 9.648$ for the
Jeffreys and uniform priors respectively, both of which satisfy
$(\mathcal{Z}_{\mathrm{L-fast}}/\mathcal{Z}_{\mathrm{L-slow}}) > 
(\mathcal{Z}_{10}/\mathcal{Z}_{00}) =8.73$. For the conservative
data, these numbers become $(\mathcal{Z}_{\mathrm{L-fast}}/
\mathcal{Z}_{\mathrm{L-slow}}) = 3.110$ and
$(\mathcal{Z}_{\mathrm{L-fast}}/\mathcal{Z}_{\mathrm{L-slow}}) = 3.137$, again
satisfying $(\mathcal{Z}_{\mathrm{L-fast}}/\mathcal{Z}_{\mathrm{L-slow}}) > 
(\mathcal{Z}_{10}/\mathcal{Z}_{00}) = 2.83$.

\begin{figure*}
\begin{center}
\includegraphics[width=18.0cm,angle=0,clip=true]{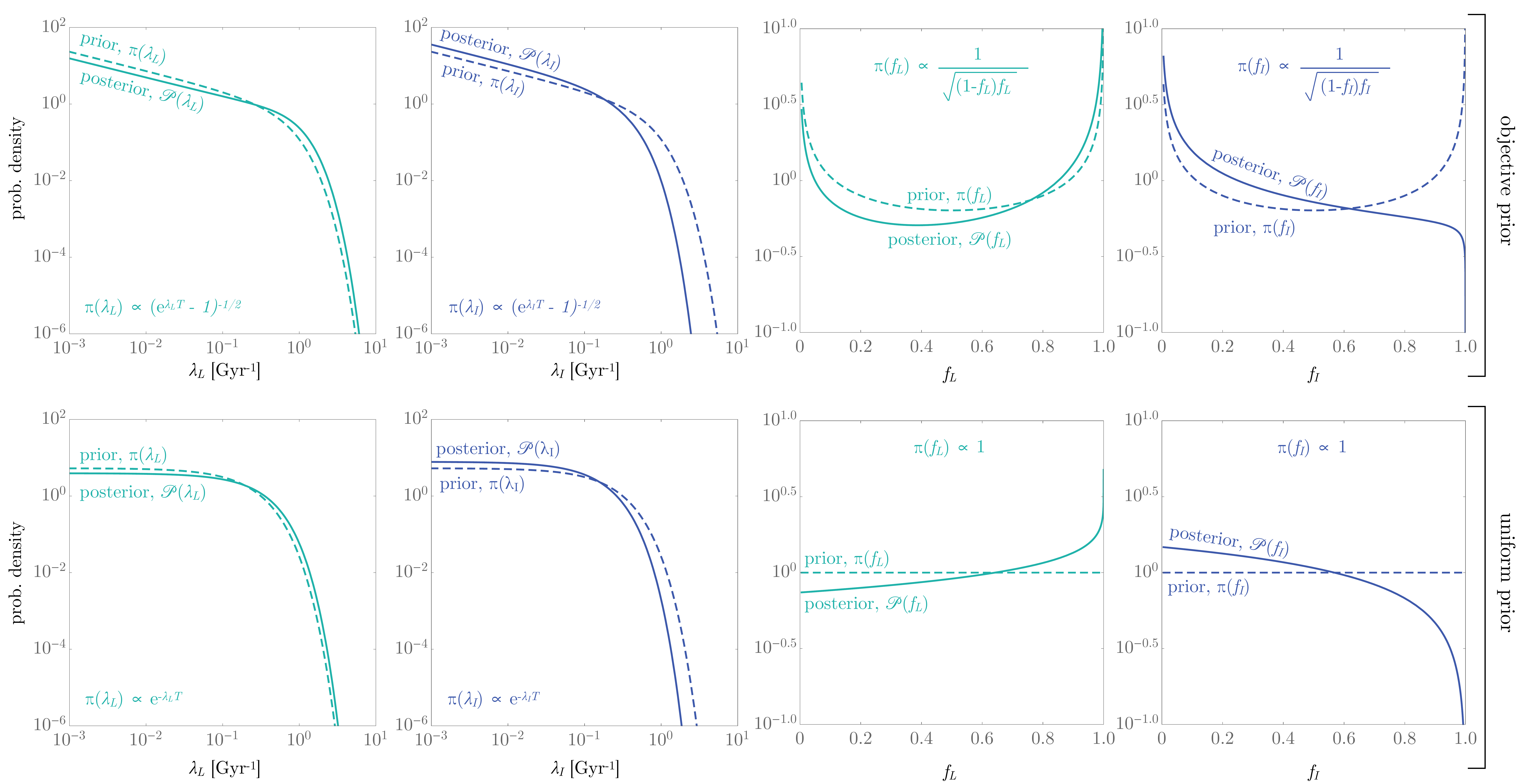}
\caption{
Top: Marginalized posteriors (solid lines) for $\lambda_L$
(far-left), $\lambda_I$ (mid-left), $f_L$ (mid-right) and $f_I$
(far-right) using the objective Bernoulli prior (dashed lines), using
the optimistic data. Lower: same as top panel but using the uniform
prior in $f$ for comparison. The $y$-axis scale is chosen to
higlight the dynamic range.
}
\label{fig:f}
\end{center}
\end{figure*}

Further understanding of these Bayes factors can be gained by evaluating the
marginalized posteriors. We numerically marginalizing the posteriors in the
case of the optimistic data and show the resulting distributions in
fig.~\ref{fig:f}. From these, one can see that the $f_L\to0$ limit drops below
the prior, whereas the $f_L\to1$ limit rises above it. Together, these results
paint a consistent picture that the timing of life's emergence, and that of
intelligent observers, favor the hypothesis that life would likely re-emerge
rapidly on the Earth were the clock to be re-ran.

\subsection*{And what of intelligence?}

Thus far, we have calculated Bayes factors concerning fast versus slow
abiogenesis rates. For the rate of emergence of intelligent observers,
Bayes factors against a fast emergence scenario tend to infinity, since
$\lim_{\lambda_I \gg 1/t_I'} \mathcal{L} \to 0$ (as shown in the SI Appendix).
Instead, it is more useful to compare the slow intelligence scenario
($\lambda_I \ll 1/t_I'$) against the ensemble of models where $\lambda_I$
can take any value, for which the Bayes factor is

\begin{align}
\frac{\mathcal{Z}_{\mathrm{I-slow}}}{\mathcal{Z}_{\mathrm{ensemble}}} &= \frac{ \int_{\lambda_L=0}^{\infty} \big(\lim_{\lambda_I \ll 1/t_I'}\mathcal{L}(\lambda_L,\lambda_I)\big)\pi(\lambda_L) \mathrm{d}\lambda_L}{
\int_{\lambda_L=0}^{\infty} \int_{\lambda_I=0}^{\infty} \mathcal{L}(\lambda_L,\lambda_I)\pi(\lambda_L) \pi(\lambda_I)\mathrm{d}\lambda_I \mathrm{d}\lambda_L
}.
\end{align}

We evaluated the above numerically using optimistic data to yield
$1.638$ and $1.474$ using the Jeffreys and uniform priors respectively.
Switching to the conservative data paints barely affects these numbers with
$1.572$ and $1.416$ (the full range of possibilities is depicted in
panel B of fig.~\ref{fig:bayesfactor}). This suggests a slight preference,
3:2 betting odds, that intelligent observers would rarely re-emerge - a
value which is broadly robust against the two priors considered and the range
of possible $t_L'$ values.

\section*{Conclusions}

In this work, we have attempted to build upon the seminal paper of
\cite{spiegel:2012} which devised a Bayesian formalism for interpreting
life's early emergence on Earth. Unlike that work, we do not treat the
timescale for life to develop into intelligent observers as a fixed quantity,
but rather infer it jointly as a free parameter. This important difference
feeds into an overall theme of the analysis presented here - to present an
objective Bayesian analysis of life's early emergence on Earth, and our
relatively late arrival within the context of Earth's habitable window.

In this vein, we have demonstrated that the commonly used power-law prior for
this problem is not objective as the results strongly depend on arbitrary
choices on the prior's domain. We show that priors on the Bernoulli probability
of life/intelligence emerging naturally provide semi-infinite support and
yield distribution which can be seen to be fair and unbiased for this problem.
Even so, it is possible to derive numerous model comparison results which are
fully independent of these priors - meaning that even those using wildly
different priors would consistently agree on the results.

The early emergence of life on Earth is naively interpreted as meaning that
if we reran the tape, life would generally reappear quickly. But if the
timescale for intelligence evolution is long, then a quick start to life is
simply a necessary byproduct of our existence - not evidence for a general
rapid abiogenesis rate. Using our objective Bayesian framework,
we show that the Bayes factor between a fast versus slow abiogenesis scenario,
is at least a factor of three - irrespective of the prior or the timescale for
intelligence evolution. This factor is boosted to nine when we replace
the earliest microfossil evidence \citep{schopf:2006,schopf:2007,schopf:2018}
with the more disputed $^{13}C$ depleted zircon deposits reported by
\cite{bell:2015}. These results are also supported by marginalizing over our
objective priors. An additional objective result concerning abiogenesis, is
that the maximum likelihood timescale for life to first appear is
$190$\,Myr after conditions became habitable ($4.21$\,Gyr ago) using the
microfossil evidence, or even just 46\,Myr using the more disputed data.
It is emphasized that these results are conditioned solely upon
the chronology data concerning life.

We find that the rate at which intelligent observers evolve is less well
constrained. Certainly, the possibility that the rate of intelligence emergence
is rapid ($\ll$Gyr) is strongly excluded, which is not surprising given that
it took several Gyr here on Earth. But the possibility that intelligence
is extremely rare and Earth ``lucked out'' remains quite viable. Overall,
we find a weak preference, 3:2 betting odds, that intelligence rarely emerges
given our late arrival.

It is tempting to apply these numbers to potentially habitable exoplanets being
discovered. However, we caution that our analysis purely concerns the Earth,
treating abiogenesis as a stochastic process against a backdrop of events and
conditions which might be plausibly unique to the Earth. If conditions
sufficiently similar to the early conditions exist and sustain on other worlds
for a Gyr or more, then our analysis would then favor the hypothesis that life
is common, by a factor of $K>3$. However, the alternative is clearly not
discounted and our Bayes factor does not cross the threshold to which it would
be conventionally described as ``strong'' ($K>10$) or ``decisive'' ($K>100$)
evidence \citep{kass:1995}. Yet, future revision regarding the earliest
evidence for life could plausibly trigger this.

Overall, our work supports an optimistic outlook for future searches for
biosignatures \citep{marais:2002,schwieterman:2018,totton:2018,fujii:2018}.
The slight preference for a rare intelligence scenario is consistent with a
straight-forward resolution to the Fermi paradox. However, our work says
nothing about the lifetime of civilizations, and indeed the weight of evidence
in favor of this scenario is sufficiently weak that searches for
technosignatures should certainly be a component in observational campaigns
seeking to resolve this grand mystery.

Data Availability Statement: All data used in this work is fully stated in the
text of this paper.

\acknow{DMK acknowledges support from the Alfred P. Sloan Foundation. Thanks
to Jason Wright for helpful comments on an early draft. Special thanks to
the anonymous reviewers for their constructive feedback.}

\showacknow{} 

\bibliography{pnas-sample}

\newpage

\onecolumn

\begin{center}
{\Huge \textbf{Supplementary Information}}
\end{center}

\section*{Full Likelihood Function and Limiting Behavior}

The main text establishes that our likelihood function is defined as
(see eqns.~\ref{eqn:bivariatepdf}-\ref{eqn:finallikefn})

\begin{align}
\mathcal{L} =& \pdf(t_L<t_L',t_I'<t_I<t_I'+t_L'|\lambda_L,\lambda_I) \nonumber\\
\qquad=& \int_{t_I=t_I'}^{t_I'+t_L'} \pdf(t_L<t_L',t_I|\lambda_L,\lambda_I) \mathrm{d}t_I,\nonumber\\
\qquad=& \int_{t_I=t_I'}^{t_I'+t_L'} \int_{t_L=0}^{t_L'} \pdf(t_L,t_I|\lambda_L,\lambda_I)\mathrm{d}t_L\mathrm{d}t_I.
\end{align}

In the main text this is evaluated to eqn.~(\ref{eqn:test1}) for the case where
$0<2t_L'+t_I'<T$. The more general result is given by

\begin{equation}
\pdf(t_L<t_L',t_I'<t_I<t_I'+t_L'|\lambda_L,\lambda_I) =
\begin{cases}
\frac{
(\lambda_L-\lambda_I) e^{-\lambda_I t_I'}
(1 - e^{-\lambda_L t_L'})
(1 - e^{-\lambda_I(T-t_L'-t_I')})
}{
\lambda_L (1-e^{-\lambda_I T}) - \lambda_I (1-e^{-\lambda_L T})
}
 & \text{if } T>2t_L+t_I ,\\
\frac{
 \lambda_L e^{-(\lambda_L-\lambda_I)t_L'-\lambda_I T}
-\lambda_I e^{(\lambda_L-\lambda_I)(t_L'+t_I')-\lambda_L T}
+(\lambda_L-\lambda_I)e^{-\lambda_I t_I'} ( 1 - e^{-\lambda_L t_L'} - e^{-\lambda_L t_I'} )
}{
\lambda_L (1-e^{-\lambda_I T}) - \lambda_I (1-e^{-\lambda_L T})
}
 & \text{if } T<2t_L+t_I .
\end{cases}
\label{eqn:like}
\end{equation}

It is useful to inspect the limiting behavior of the likelihood function, which
are depicted in the side panels of fig.~\ref{fig:likelihood} of the main text. In the limit of fast
abiogenesis, $\lambda_L \gg 1/t_L'$, one finds that

\begin{align}
\lim_{\lambda_L \gg 1/t_L'} \mathcal{L} &=
\frac{ e^{-\lambda_I t_I'} - e^{-\lambda_I (t_L'-t_I')} }{ 1 - e^{-\lambda_I T} },
\end{align}

whereas in the slow limit we find

\begin{equation}
\lim_{\lambda_L \ll 1/t_L'} \mathcal{L} =
\begin{cases}
\frac{
\lambda_I t_L' e^{-\lambda_I t_I'} (1-e^{-\lambda_I t_L'})
}{
\lambda_I T - (1 -e^{-\lambda_I T})
}
 & \text{if } T>2t_L+t_I ,\\
\frac{
e^{-\lambda_I t_I'} [\lambda_I t_L' + e^{-\lambda_I t_L'} (1-\lambda_L(T-t_I'-t_L'))] - e^{-\lambda_I(T-t_L')}
}{
\lambda_I T - (1 -e^{-\lambda_I T})
}
 & \text{if } T<2t_L+t_I .
\end{cases}
\end{equation}

Similarly, one can define limits for slow and fast intelligence evolution as

\begin{align}
\lim_{\lambda_I \gg 1/t_I'} \mathcal{L} &= 0,
\end{align}

and

\begin{equation}
\lim_{\lambda_I \ll 1/t_I'} \mathcal{L} =
\begin{cases}
\frac{
\lambda_L t_L' (1-e^{-\lambda_L t_L'})
}{
\lambda_L T - (1 -e^{-\lambda_L T})
}
 & \text{if } T>2t_L+t_I ,\\
\frac{
\lambda_L t_L' + e^{-\lambda_L t_L'}[1-\lambda_L(T-t_I'-t_L')] - e^{-\lambda_L(T-t_I'-t_L')}
}{
\lambda_L T - (1 -e^{-\lambda_L T})
}
 & \text{if } T<2t_L+t_I .
\end{cases}
\end{equation}

Finally, one can evaluate the likelihood function in the corners of the entire
parameter volume taking limits of these results another time. Since
$\lim_{\lambda_I \gg 1/t_I'} \mathcal{L} = 0$, then it makes no difference what
value $\lambda_L$ takes in along the $\lambda_I \gg 1/t_I'$ axis, the result is
still zero. For the $\lim_{\lambda_I \ll 1/t_I'} \mathcal{L}$ case, we have
two non-zero corners of

\begin{equation}
\lim_{\lambda_L \ll 1/t_L'} \lim_{\lambda_I \ll 1/t_I'} \mathcal{L} =
\begin{cases}
\frac{
2 t_L'^2
}{
T^2
}
 & \text{if } T>2t_L+t_I ,\\
\frac{
2T (t_I'+2t_L') - T^2 - t_I^2 - 2t_L'^2 - 4t_L't_I'
}{
T^2
}
 & \text{if } T<2t_L+t_I,
\end{cases}
\end{equation}

and

\begin{align}
\lim_{\lambda_L \gg 1/t_L'} \lim_{\lambda_I \ll 1/t_I'} \mathcal{L} = \frac{2t_L'}{T}.
\end{align}

\section*{Max Likelihoods}

\begin{figure}
\begin{center}
\includegraphics[width=16.0cm,angle=0,clip=true]{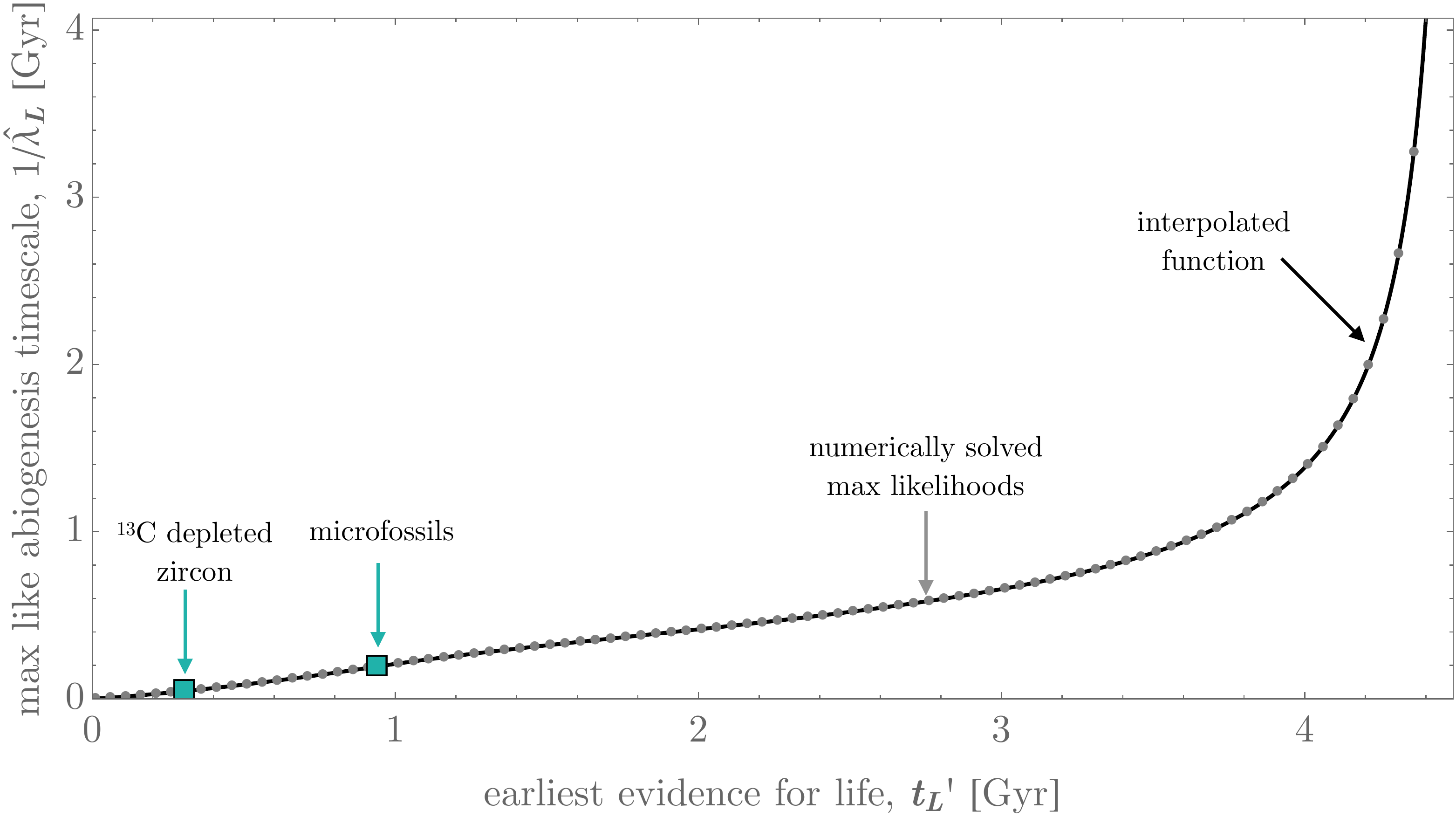}
\caption{
Maximum likelihood value of $1/\lambda_L$ as a function of $t_L'$ (the
earliest evidence for life).
}
\label{fig:maxlike}
\end{center}
\end{figure}

It is straight-forward to maximize the likelihood function as a function of the
parameters $\lambda_L$ and $\lambda_I$. The maximum likelihood always lies along
the $\lambda_I \ll 1/t_I'$ axis, since the likelihood function monotonically
decreases as $\lambda_I$ increases. We therefore simply need to maximize the
function $\lim_{\lambda_I \ll 1/t_I'} \mathcal{L}$, found earlier. Rather than
do this exclusively fr the optimistic and conservative data, we repeat the 
calculation along a sliding scale to illustrate the full behavior (see
fig.~\ref{fig:maxlike}).

\section*{Imposing Fairness on the $\lambda$ Power-Law Prior}

Starting from eqn.~(\ref{eqn:powerlaw}) of the main text, we can transform the prior into 
$f$, where $f \equiv 1 - e^{-\lambda T}$ using

\begin{align}
\pdf(\lambda) \Big(\frac{\mathrm{d}\lambda}{\mathrm{d}f}\Big) \mathrm{d}f &= \pdf(f) \mathrm{d}f
\end{align}

and noting that $\lambda = -T^{-1} \log(1-f)$, yielding

\begin{equation}
\pdf(f) =
\begin{cases}
\frac{1}{\log(-\log(1-f_{\mathrm{max}})) - \log(-\log(1-f_{\mathrm{min}}))}
\frac{1}{(1-f) \log(-\log(1-f))}
 & \text{if } n=-1 ,\\
\frac{n+1}{1-f} \frac{(-\log(1-f))^n}{ (-\log(1-f_{\mathrm{max}}))^{n+1} - (-\log(1-f_{\mathrm{min}}))^{n+1} }. & \text{otherwise },
\end{cases}
\label{eqn:fpower}
\end{equation}

for which the more general case of $n \neq 1$ is given in eqn.~(\ref{eqn:9}) of the main
text. In the above, the limits can be related as

\begin{align}
f_{\mathrm{min}} = 1 - e^{ -\lambda_{\mathrm{min}} T },\nonumber\\
f_{\mathrm{max}} = 1 - e^{ -\lambda_{\mathrm{max}} T }.
\end{align}

Let us now consider whether this prior is biased with respect to $f$ by
evaluating the odds ratio of obtaining an a-priori low value of $f$ ($f<0.5$)
versus a high value ($f>0.5$) using the definition of $\mathbb{F}$ given by
eqn.~(\ref{eqn:Fdefinition}) of the main text. Evaluating this function yields

\begin{equation}
\mathbb{F} =
\begin{cases}
\frac{ \log\big( -\frac{\log(1-f_{\mathrm{max}})}{\log2} \big) }{\log\big( -\frac{\log2}{\log(1-f_{\mathrm{min}})} \big)}
 & \text{if } n=-1 ,\\
\frac{ (\log2)^{n+1} - (-\log(1-f_{\mathrm{max}}))^{n+1} }{ -(\log2)^{n+1} + (-\log(1-f_{\mathrm{min}}))^{n+1} }
. & \text{otherwise },
\end{cases}
\end{equation}

Let's define $\mathcal{F} \equiv \log(1-f)$ (such that $\lambda =
-\mathcal{F}/T$) and then set the above equal to unity and solve for
$\mathcal{F}_{\mathrm{max}}$ to give

\begin{equation}
\lim_{\mathbb{F} \to 1} \mathcal{F}_{\mathrm{max}} =
\begin{cases}
\frac{ (\log2)^2 }{ \mathcal{F}_{\mathrm{min}} }
 & \text{if } n=-1 ,\\
-\Big( (2\log2)^{n+1} + \mathcal{F}_{\mathrm{min}} (-\mathcal{F}_{\mathrm{min}})^n \Big)^{1/(n+1)}
. & \text{otherwise },
\end{cases}
\label{eqn:fairFmax}
\end{equation}

And this may now be easily related back to $\lambda$ as

\begin{equation}
\lim_{\mathbb{F}_f \to 1} \lambda_{\mathrm{max}} =
\begin{cases}
\frac{ (\log2)^2 }{ T^2 \lambda_{\mathrm{min}} }
 & \text{if } n=-1 ,\\
T^{-1} \Big( (2\log2)^{n+1} - ( \lambda_{\mathrm{min}} T )^{n+1} \Big)^{1/(n+1)}
. & \text{otherwise },
\end{cases}
\label{eqn:ideallambdamax}
\end{equation}

Note, in order for $\lambda_{\mathrm{max}}>\lambda_{\mathrm{min}}$,
this requires $\lambda_{\mathrm{min}}<T^{-1}\log2 \sim 0.14$\,Gyr$^{-1}$.

\section*{Bias in the $\lambda$ Power-Law Prior}

\begin{figure}
\begin{center}
\includegraphics[width=16.0cm,angle=0,clip=true]{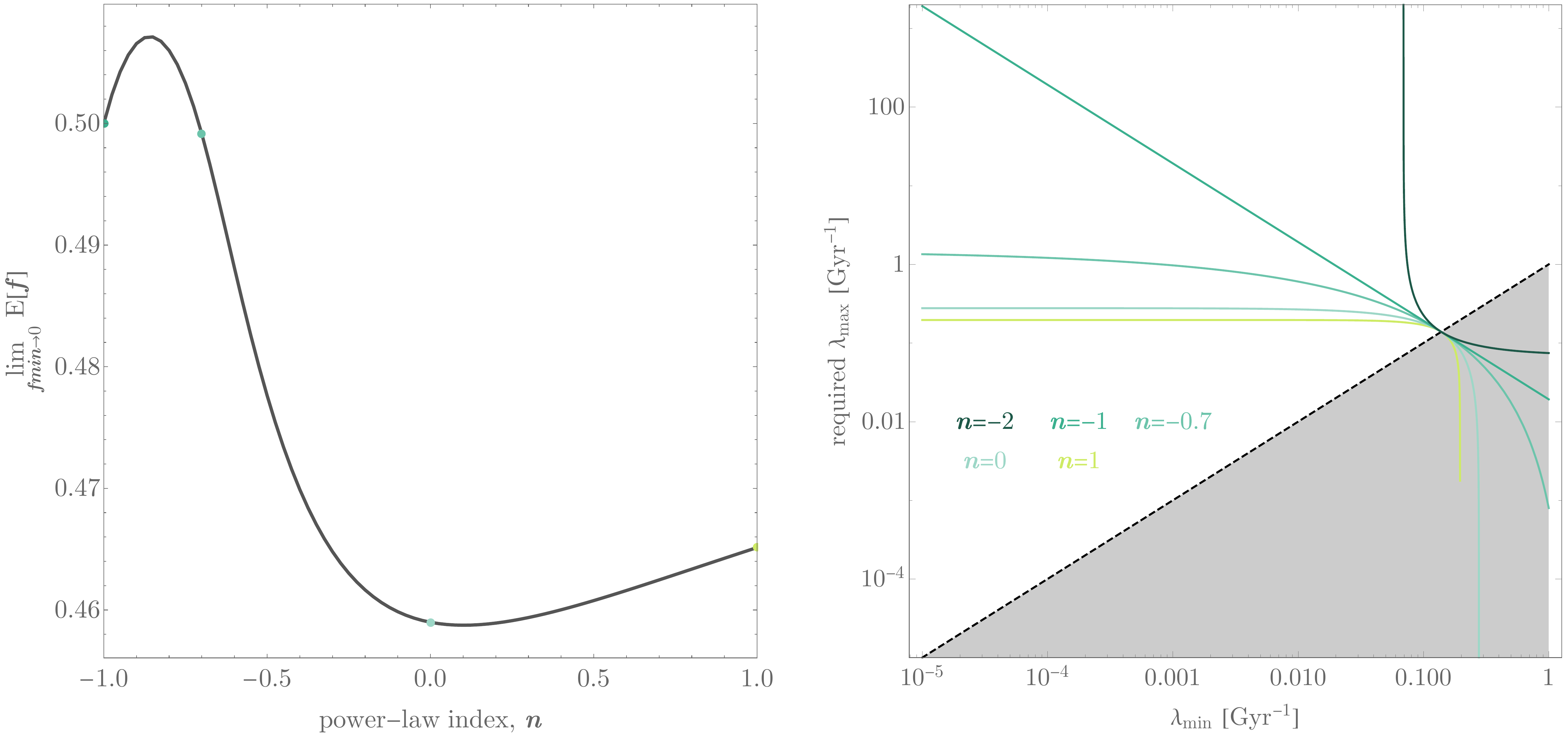}
\caption{
Left: The \textit{a-priori} expectation value of $f$ (the probability of at
least one success in a Poisson process) for priors of the form $\lambda^n$
when imposing a fairness using eqn.~(\ref{eqn:fairFmax}). The expectation
values shown are all in the limit of $f_{\mathrm{min}} \to 0$ and we
find that the $n=-1$ (log-uniform prior) naturally recovers an unbiased
result of one-half. Right: Required $\lambda_{\mathrm{max}}$ value to impose
a fair prior as a function of $\lambda_{\mathrm{max}}$ for several highlighted
choices of $n$. Only $n=-1$ provides the necessary dynamic range to handle
the scope of our problem.
}
\label{fig:expf}
\end{center}
\end{figure}

The \textit{a-priori} expectation value of $f$ should be equal to one half for
an unbiaed prior. This may be evaluated through integration of 
eqn.~(\ref{eqn:fpower}), yielding

\begin{equation}
\mathrm{E}[f] =
\begin{cases}
\frac{
  (n+1) \Gamma[n+1,-\mathcal{F}_{\mathrm{min}}]
- (n+1) \Gamma[n+1,-\mathcal{F}_{\mathrm{max}}]
-(-\mathcal{F}_{\mathrm{max}})^{n+1}
}{
(-\mathcal{F}_{\mathrm{min}})^{n+1} - (-\mathcal{F}_{\mathrm{max}})^{n+1}
}
 & \text{if } n=-1 ,\\
\frac{
\mathrm{Ei}[\mathcal{F}_{\mathrm{min}}] - \mathrm{Ei}[\mathcal{F}_{\mathrm{max}}]
+\log(\mathcal{F}_{\mathrm{max}})-\log(\mathcal{F}_{\mathrm{min}})
}{
\log(-\mathcal{F}_{\mathrm{max}}) - \log(-\mathcal{F}_{\mathrm{min}})
}
& \text{otherwise },
\end{cases}
\label{eqn:fexpectation}
\end{equation}

where $\mathrm{Ei}[x]$ is the exponential integral function. We may now
replace $\mathcal{F}_{\mathrm{max}}$ using eqn.~(\ref{eqn:fairFmax}), to impose
a fair prior. After doing so, we take the limit of the expectation value
as $f_{\mathrm{min}} \to 0$ and show the result as a function of $n$
in fig.~\ref{fig:expf}. This reveals that the $n=-1$ prior converges to
precisely one-half as desired, whereas other indices (in general) do not.

Aside from $n=-1$ being unbiased, we also see that $n \simeq -0.709$ appears 
unbiased prior too. This can be understood by the fact that the minimum
in $\pdf(f)$ occurs at exactly 1/2 when $n=-\log2=-0.693...$. However,
$\pdf(f)$ is not symmetric about $f=1/2$ when using $n=-\log2$ and thus
is not a fair prior. Accordingly, a small perturbation away from $n=-\log2$
is needed to obtain an unbiased distribution.

Unfortunately, it is not practical to use $n=-0.709$ as an alternative
fair and unbiased prior. This can be understood by considering the behavior
of eqn.~(\ref{eqn:ideallambdamax}) in the limit of
$\lambda_{\mathrm{min}}\to0$. Unless, $n=-1$, the limit is
$2^{1/(n+1)} T^{-1} \log2$, which in general is not large enough to serve
as a useful upper limit on the prior. For example, with $T=5$\,Gyr and
$n=-0.709$, this yields $\lambda_{\mathrm{max}} = 1.5$\,Gyr$^{-1}$ 
- which means that the posterior can never explore solutions with more than 1.5
successes per billion years. On that basis, it is argued here that this is an
overly restrictive choice of prior distribution that significantly truncates
the posterior. And thus, even though an $n=-0.709$ power-law
prior can be fair and unbiased, it will not be considered further.

\section*{An Alternative Explanation of Prior Independent Bayes Factors}

In the main text, it is shown that for the four models which define the four 
corners of the two-dimensional parameter volume, objective Bayes factors can
be expressed independent of the prior $\pi(\lambda_L,\lambda_I)$. We here
provide an alternative explanation of this interesting feature using the
Savage-Dickey ratio instead.

This can be accomplished by first considering that the extreme model
$\mathbf{\mathcal{M}_{00}}$ is embedded within the nested model of arbitrary
$\{\lambda_L,\lambda_I\}$. If we compare the extreme (and nested) model to the
more general case, the Bayes factor will be given by the Savage-Dickey
ratio \citep{dickey:1971}:

\begin{align}
\frac{\mathcal{Z}_{00}}{\mathcal{Z}_{\mathrm{ensemble}}} &= \lim_{\lambda_L \ll 1/t_L'} \lim_{\lambda_I \ll 1/t_I'} \Big(\frac{ \mathcal{P}(\lambda_L,\lambda_I) }{ \pi(\lambda_L,\lambda_I) }\Big),
\end{align}

where $\pi$ is the prior, $\mathcal{P}$ is the posterior and we use the
notation ``00'' to denote model $\mathbf{\mathcal{M}_{00}}$  and ``ensemble''
to denote the general case comprising the ensemble of scenarios. Note that the
posterior density featured above is not marginalized, but rather directly
given by the closed form formula of our likelihood multiplied by the prior.
Accordingly, prior will cancel out leaving

\begin{align}
\frac{\mathcal{Z}_{00}}{\mathcal{Z}_{\mathrm{ensemble}}} &= \lim_{\lambda_L \ll 1/t_L'} \lim_{\lambda_I \ll 1/t_I'} \mathcal{L}(\lambda_L,\lambda_I),
\end{align}

thus providing a totally prior-free measure of the Bayes factor. To finish
the analysis, one may now compare different extrema against one another
noting that the arbitrary ``ensemble'' models will cancel out. For example,
$(\mathcal{Z}_{10}/\mathcal{Z}_{00}) =
(\mathcal{Z}_{10}/\mathcal{Z}_{\mathrm{ensemble}})/(\mathcal{Z}_{00}/\mathcal{Z}_{\mathrm{ensemble}})$.

\end{document}